# Integrated Metasurface-based Wavelengths Division Demultiplexers

**Amged Alquliah,[1,2] Mohamed ElKabbash,[4,\*] JinLuo Cheng,[1,2] Wei Li,[1,2] and Chunlei Guo[3,\*]**

[1]GPL Photonics Laboratory, State Key Laboratory of Applied Optics, Changchun Institute of Optics, Fine Mechanics and Physics, Chinese Academy of Sciences, Changchun, China
[2]University of Chinese Academy of Sciences, Beijing 100039, China
[3]The Institute of Optics, University of Rochester, Rochester, New York 14627, USA
[4]The Research Laboratory of Electronics, Massachusetts Institute of Technology, Cambridge, MA 02139, USA
[\*]Corresponding authors: melkabba@mit.edu, and guo@optics.rochester.edu

**ABSTRACT:** We present a design approach for realizing on-chip wavelength division demultiplexing (WDD) schemes by integrating all-dielectric metasurfaces of $TiO_2$ nanorod arrays into a SiN waveguide. The designed metasurface locally modifies the effective refractive index of the SiN waveguide, creating an effective WDD that selectively passes a certain band of wavelengths into a specific output port. A set of representative 2-channel and 3-channel WDDs schemes were demonstrated for input $TE_{00}/TM_{00}$ modes and operating in different bands, showing the flexibility of our design approach. The proposed WDD schemes are compatible with visible to infrared wavelengths, photolithography-based fabrication, high efficiency with maximum transmission of 91%, and a small footprint at a few microns. Our design method paves the way for realizing several on-chip integrated devices for applications in optical data processing and biological sensing.

## 1.1 Introduction

The massive growth of computational load by data-intensive applications such as artificial intelligence and machine learning has necessitated increasing the capacity of computing systems. Photonic Integrated Circuits (PICs), unlike their electronic counterpart, provide higher bandwidth density, lower power consumption, and low-loss data transport [1, 2]. Recently PICs have poised to play an essential role in a range of applications including optical information processing [3-6], neuro-inspired computing [7-10], quantum computing [11, 12], and microwave photonics [13, 14]. Wavelengths division demultiplexing (WDD), which separates and routes different wavelengths into distinct channels, is a key functionality in PICs [15, 16]. This technique multiplies the bandwidth density of a single photonic waveguide by utilizing different wavelengths as unrelated carrier signals [17-19].

Several approaches are used to realize WDDs, such as échelle grating [20, 21], arrayed waveguide gratings [22-24], arrayed ring resonators [25, 26], photonic crystal [27, 28], and Multimode/Mach-Zehnder interferometers [29-32]. However, these approaches suffer from having relatively large footprints (10s-100s µm) that limits their large-scale integration [15]. Plasmonic-WDDs show more compact footprints, but they suffer from significant optical losses [33-37]. In contrast, WDDs-based microcavities exhibit high efficiency but narrow operating bandwidth and strong sensitivity to temperature, fabrication imperfections, and the surrounding environment [38, 39]. Devices employing the inverse design approach exhibit ultracompact footprints and efficient performance [15, 16, 40, 41]. However, they are limited to simple photonic structures [7, 42, 43], and the fabrication of inverse-design-based WDDs may not be fully compatible with photolithography which is necessary for large-scale production [10, 16, 44]. Recently, genetic algorithms and finite element methods have been used to realize ultrasmall WDDs [45]. But in most of the proposed structures based on the intelligent algorithm, the insertion losses were significantly high and the crosstalk between the output channels was not considered.

On the other hand, the advent of all-dielectric meta-photonics, which denotes structures that control the propagation of light at the nanoscale by arrays of ultrathin all-dielectric nanoantennas [46-48] has opened up a new research frontier in photonics[49]. Adopting this design concept, numerous compact, highly efficient, and low-loss photonic devices have been recently reported for free space [50-55] and guided-wave applications [56-62].[49, 60-62]

Here, we report a design approach for realizing on-chip wavelength-division demultiplexing devices by integrating an all-dielectric metasurface consisting of $TiO_2$ nanorod-arrays into a SiN waveguide. Our design relies on the wavelength-dependent local modification of the effective refractive index of the SiN waveguide induced by arrays of $TiO_2$ nanoantennas. To prove the feasibility of our design approach, we propose various broadband WDDs devices that are compatible with infrared wavelengths (important for telecom applications) and with near-visible/visible wavelengths which are crucial to quantum computing and optogenetics. Moreover, we extend our approach from two to three-channel WDDs. The proposed WDDs devices operate for input $TE_{00}$ and $TM_{00}$ modes. These WDDs devices are efficient (maximum transmission 91 %), compact (a footprint of only a few microns), and compatible with photolithography-based fabrication. Our design method paves the way for realizing several on-chip integrated devices such as mode-selective polarization demultiplexer and hybrid mode/wavelength demultiplexer.

## 1.2 Materials and design approach

The materials are selected to realize WDDs operating at different spectral regions with minimal absorption losses. SiN was chosen as a waveguiding medium over silicon because it provides a wider transparency window ($\lambda$=0.25 µm - 8 µm), lower propagation losses, and higher fabrication flexibility [63, 64]. Amorphous-TiO$_2$ (a-TiO$_2$) was selected for the nanoantenna arrays because of the following reasons: first, a-TiO$_2$ enjoys a wide transparency window [65, 66]; second, a-TiO$_2$ has a higher refractive index than SiN, which increases the confinement of guided light inside the nanoantenna and thus enables engineering the effective medium of the SiN waveguide. The complex optical constants of SiN and a-TiO$_2$ were obtained from the Palik database [67].

The design flowchart for a *n*-channel WDD (with *n*=2 or 3) is shown in **Fig. 5-1a**. We first assign the number of *n*. After that, we place *n* nanoantennas-arrays as metasurface. Then, we calculate the effective index ($n_{eff}$) of the $TE_{00}$ or $TM_{00}$ modes in the hybrid waveguide (i.e., nanoantenna-loaded SiN waveguide) at the target wavelengths using the FDE solver (Ansys, Lumerical Inc.). Then, we adjust the period of the placed nanoantennas-arrays to allow passing certain wavelength bands to the predefined output channels of the WDD. Finally, we optimize the device performance metrics (i.e., transmission and output channels crosstalk) by fine-tuning the footprint and filling ratio of nanoantennas-arrays. The device performance was verified using 3D FDTD (Ansys, Lumerical. Inc). The designed structure was oriented along the x-direction, and the simulation set-up was enclosed by PML boundary conditions. A non-uniform auto mesh setting was applied in the simulation region with a (2.5 nm) conformal mesh size. The $TE_{00}$/$TM_{00}$ mode was injected into the input port using a mode source and with enabling multi-frequency calculation (>>100 frequency points). The frequency-domain time power monitors were utilized to record the transmission and field intensity profiles in the devices. **Figure 1b** shows the calculated effective refractive index of TE$_{00}$ and TM$_{00}$ modes in the bare SiN waveguide and in a hybrid waveguide over the simulated wavelengths. As it can be seen. the $n_{eff}$ of TE$_{00}$/TM$_{00}$ mode in the hybrid waveguide is higher than that in the bare waveguide. Therefore, the effective medium of the SiN waveguide can be locally engineered by fine-tuning the parameters and spatial patterning of a-TiO$_2$ nanoantennas-arrays located atop the SiN waveguide. **Figure 5-1c** shows the electric field components of input TE$_{00}$ and TM$_{00}$ in a bare SiN waveguide.

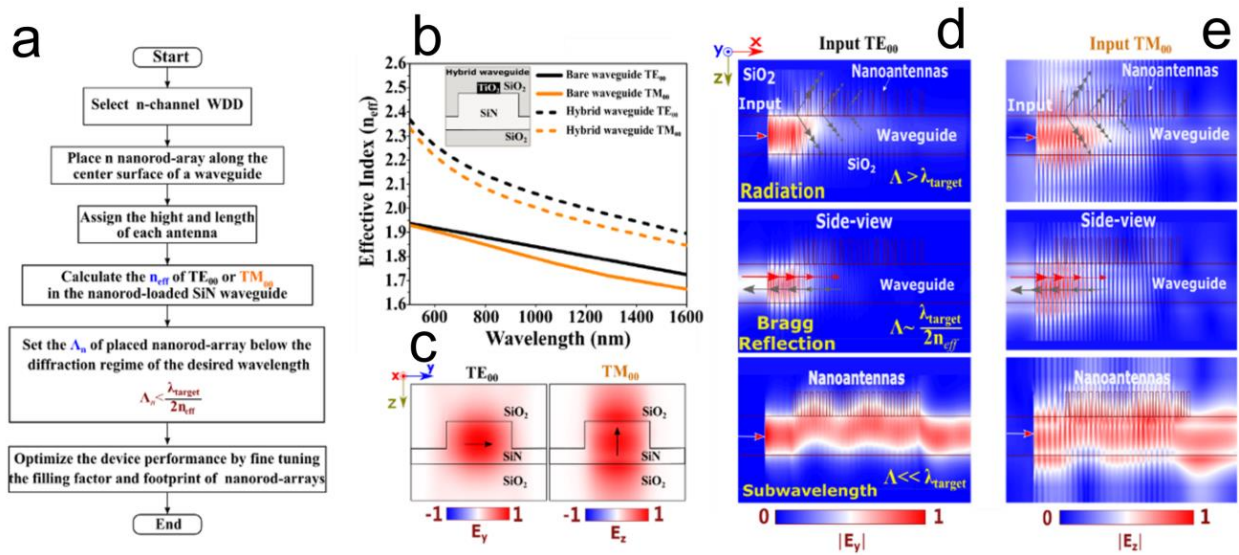

**Figure 1:** Design approach and mechanism of proposed metasurface-based integrated wavelength demultiplexer. **a)** Flowchart for the proposed design, where $n_{eff}$ is the effective index of the TE$_{00}$/TM$_{00}$ mode in TiO$_2$ nanoantennas -loaded SiN waveguide, $\Lambda_n$ is the period of *n* nanoantennas-arrays, and $\lambda_{target}$ is the wavelength of injected light. **b)** Effective refractive index of the fundamental TE and TM Modes in a bare SiN waveguide and in a hybrid nanoantenna/waveguide over the simulated wavelengths. The inset figure shows a cross-section view of the hybrid waveguide. **c)** Profiles of the fundamental TE and TM modes in a bare SiN waveguide at 1550 nm. Arrows indicate the electric field components of the corresponding modes. **d-e)** Side-views (cutting planes (XZ)) show the spatial distribution of light propagation along the x-axis in the device at 1550 nm, as a function of wavelength-to-period ratio in three distinct operating regimes: (top) radiation, where ($\Lambda > \lambda_{target}$.); (middle) Bragg reflection, where ($\Lambda \approx \lambda_{target}/2n_{eff}$.); and (below) sub-wavelength, where $\Lambda \ll \lambda_{target}$. These three regimes were employed for our design to allow passing certain wavelengths to predefined outputs, thus realizing integrated WDDs. The solid black lines and rectangles indicate the boundaries of the SiN waveguide and TiO$_2$ nanoantennas.

The operating principle of the design relies on engineering the effective medium of the SiN waveguide for realizing on-chip WDDs that selectively passes a certain wavelength band to specific output ports. This can be implemented by tuning the wavelength-to-pitch ratio

($\bar{\lambda} = \lambda_{target}/\Lambda$) of the a-TiO$_2$ nanorod-arrays [68-70], $\Lambda$ is the period of nanoantennas-arrays.

**Figures 1d and Figure 1e** show simulated field intensity ($\lambda_{target} = 1550\ nm$) at the XZ plane when the TE$_{00}$ and TM$_{00}$ are injected in the structure (x-axis) from left to right. As it can be seen, for a given $\lambda_{target}$, the device operates at three different regimes depending on the period of the nanorod-arrays [68, 70]. These operation regimes are as follows: Radiation $\Lambda > \lambda_{target}$, where the light scatters out of the SiN waveguide; Bragg reflection ($\Lambda \approx \lambda_{target}/2n_{eff}$), which corresponds to the photonic bandgap where light cannot pass through the nanorod-arrays [68, 70]; and sub-wavelength ($\Lambda \ll \lambda_{target}$), where structure behaves as a homogenous waveguide thus allows light to propagate without diffraction. Metasurfaces are subwavelength elements that provide local control over the phase of light. In the context of guided propagation, metasurfaces have shown an exceptional ability to yield highly efficient and broadband photonic integrated devices with unrivaled compactness [56, 57, 59, 71, 72]. Note that, since the nanorod-arrays are periodic in a longitudinal direction (x-direction), but not in the transversal z and y directions, numerical analysis employing the supercell lattice methods are not suitable, and optimized numerical tools should be used [68, 70, 73]. To ease the optimization procedures of such multi-dimensional structures and simplify the physical mechanism of such structures, the effective medium theory is typically employed. For each subwavelength nanorod-arrays, the equivalent medium effective index is approximated by Rytov's equation as follows [74]:

$$n_{n\_array\ \|}^2 = \frac{W_n}{\Lambda_n} n_{hybrid}^2 + (1 + \frac{W_n}{\Lambda_n}) n_{bare}^2 \qquad 1$$

where $n_{n\_array\ \|}^2$, $n_{hybrid}^2$ and $n_{bare}^2$ are the equivalent medium effective index for light propagation through the nanoantenna-array (along the x-direction), effective indices of hybrid waveguide (i.e., nanoantenna-loaded waveguide) and bare SiN waveguide, respectively, and $\frac{W_n}{\Lambda_n}$ is the filling ratio of n nanorod-arrays.

## 1.3 Results and discussion

### 1.3.1 Two-channel WDDs

The functionality and schematic layout of the two-channel WDDs are illustrated in **Fig. 2a**. The structure comprises a TiO$_2$ metasurface (black rods) superimposed on a center surface of a ridge SiN waveguide (input port) that has a width ($d$=3.5 µm), a ridge height ($h$=220 nm) and under-etched film height ($s$=200 nm) (**Fig. 2b**). The stem SiN waveguide (input port) is then connected to two output branches (output ports) having a width of (1.75 µm). The SiN input waveguides are bonded on a 2-µm layer of Silica substrate.

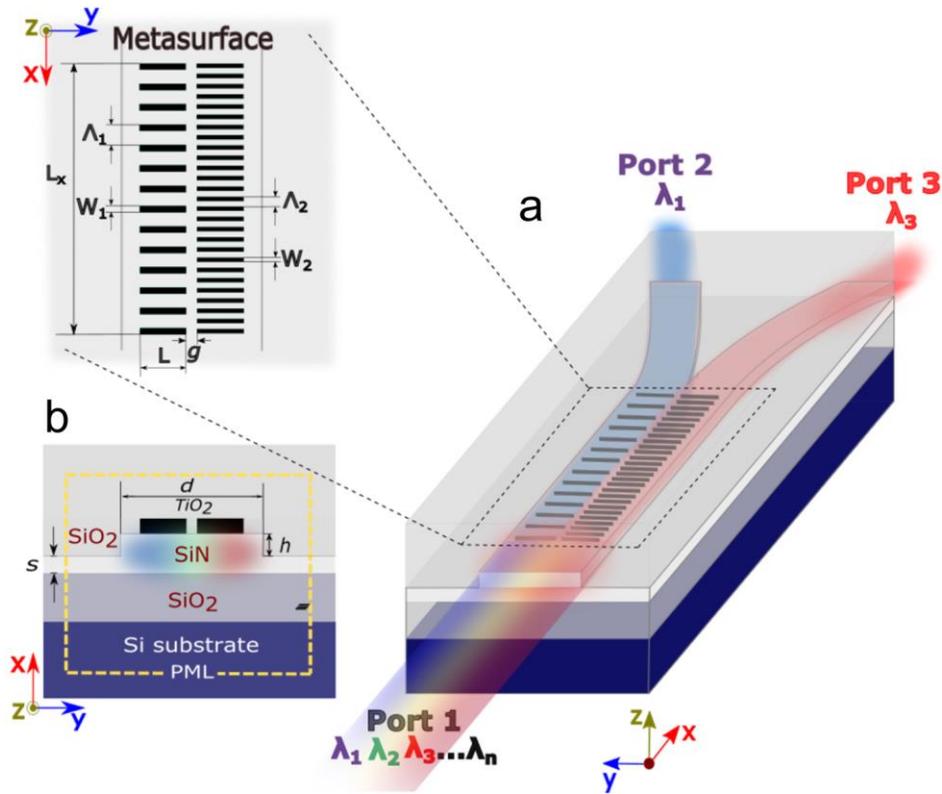

**Figure 2:** Design of two-channel WDDs. **a)** The schematic illustration of the proposed metasurface-based integrated 2-channel WDDs. Inset (top left) shows the top view and parameters of TiO2 nanoantenna arrays located on the top surface of the SiN waveguide. **b)** Cross-sectional view shows the materials system used in the device and the simulation set-up (dashed green square). Note that the dimensions of input and output waveguides are fixed for all designed two-channel WDDs devices.

The device is further supported on a Silicon substrate and over-coated with a 3-μm silica layer to increase the modal confinement inside the waveguides [56]. The a-TiO$_2$ metasurface consists of two nanoantenna-arrays of a rectangular cross-section with identical length ($L$=1.1 μm), height ($h$=250 nm), and footprint ($L_x$=10 μm). The arrays are separated by a gap ($g$=180 nm) and oriented along the x-axis. Each array consists of $n$ nanorods, width $W_n$, and pitch period $\Lambda_n$ (inset **Fig. 2a**). Supplementary **Table S1** provides detailed parameters for the optimized nanoantenna-arrays of the proposed two-channel WDDs devices. The dependence between the target wavelengths of the WDDs devices and the pitch period of the antenna-array ($n$) is shown in **Figs. S1**. Note that, the refractive index contrast $\Delta n$ between the SiN waveguide and dielectric TiO$_2$ nanorod-arrays over the simulated region strengthens the interaction between the TE$_{00}$/TM$_{00}$ modes in the waveguide and the leaky modes in dielectric a-TiO$_2$ nanorods (Appendix 3 **Fig. S2**). The equivalent medium indices $n^2_{n\_array\,\|}$ of the designed nanorod-arrays as a function of its filling factor (i.e., defined as $W_n/\Lambda_n$) is illustrated in **Fig. S3**.

### a) Device performance for TE$_{00}$ mode at different spectral regions.

**Figure 3a, Fig. 3c and Fig. 3e** show the field intensity profiles $|E|^2$ for an input TE$_{00}$ mode along the propagation direction (x-axis) of the device for a wavelength range of infrared region 1280 nm-1050 nm (a), near-visible region 1000-850 nm (c), **and** visible region 640 nm-560 nm (e). The inset depicts the out-of-plane spatial distribution of field intensity $|E|^2$ through the center of TiO$_2$ nanorods at the XY plane. The field intensity $|E|^2$ changes from the top side of the TiO$_2$ metasurface (top nanorod-array) to the lower, showing the ability of our designed metasurface to selectively allow a certain wavelength band to propagate through a specific nanorod-arrays. Following the interaction with the metasurface, the input mode in the target wavelength band takes a relatively confined path in its predefined output. The proposed approach is suitable for realizing efficient WDDs for wavelengths at different spectral regions from visible

to infrared. **Figure 3b, 3d and 3f** show the calculated transmission at output ports (port2 and port3) of the respective WDDs devices. Our devices enjoy broadband demultiplexing performances of wide bands of wavelengths (i.e., large channel spacing).

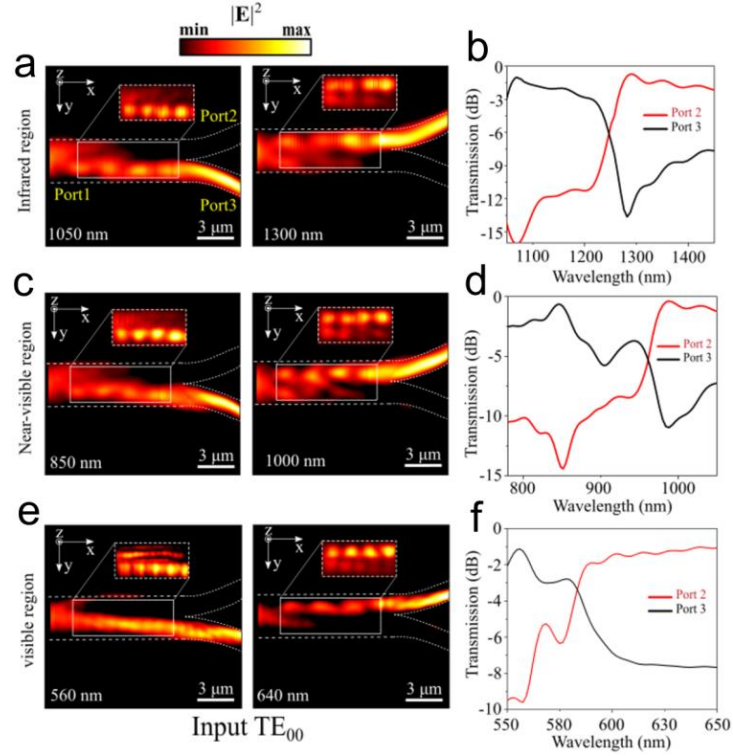

**Figure 3.** WDDs performances for input $TE_{00}$ mode. **a, c, e)** The simulated field intensity profiles $|E|^2$ along the proposed WDDs at the (XY) plane for the fundamental $TE_{00}$ mode operating in the infrared, near-visible and visible regions, respectively. The dashed lines and rectangles in the figures show delineate the boundaries of the input/output SiN ports and the superimposed $TiO_2$ metasurface, respectively. A clear spatial separation for the different wavelengths' bands can be observed. The insets show the out-of-plane field intensity $|E|^2$ distribution through the center of the metasurface. The fields are recorded at the central wavelengths of the bands. **b, d, f)** The calculated transmission at the output ports of the corresponding WDDs devices over the simulated wavelengths. The pass-bands are clearly distinguished in these figures. The observed transmission peaks are at 1050 nm, 1280 nm, 850, 1000 nm, 560 nm and 640 nm, respectively.

The peak transmissions are -1 dB in 1050 nm band, -1.2 dB in 1280 nm band, -0.84 dB in 850 nm band, -0.98 dB in 1000 nm band, -1.1 dB in 560 nm band and -1.02 dB in 640 nm band, respectively. The lowest simulated crosstalk in the respective bands is under -15.7 dB. Our simulations indicate that the losses in the device are mainly due to the reflection and scattering events caused by the strong light-antennas interaction as well as the impedance mismatch by

TiO$_2$ nanorods.

**b) Device performance for TM$_{00}$ mode at different spectral regions.**

Similar to the results in **Figure 3, Figure 4a, Fig. 4c, and Fig. 4e** show the simulated field intensity profiles $|E|^2$ for input TM$_{00}$. **Figure 4b, Fig. 4d, and Fig. 4f** show the calculated transmission at output ports (port2 and port3) of the respective WDDs devices operating for TM$_{00}$ mode. The peak transmission is -1 dB in the 1100

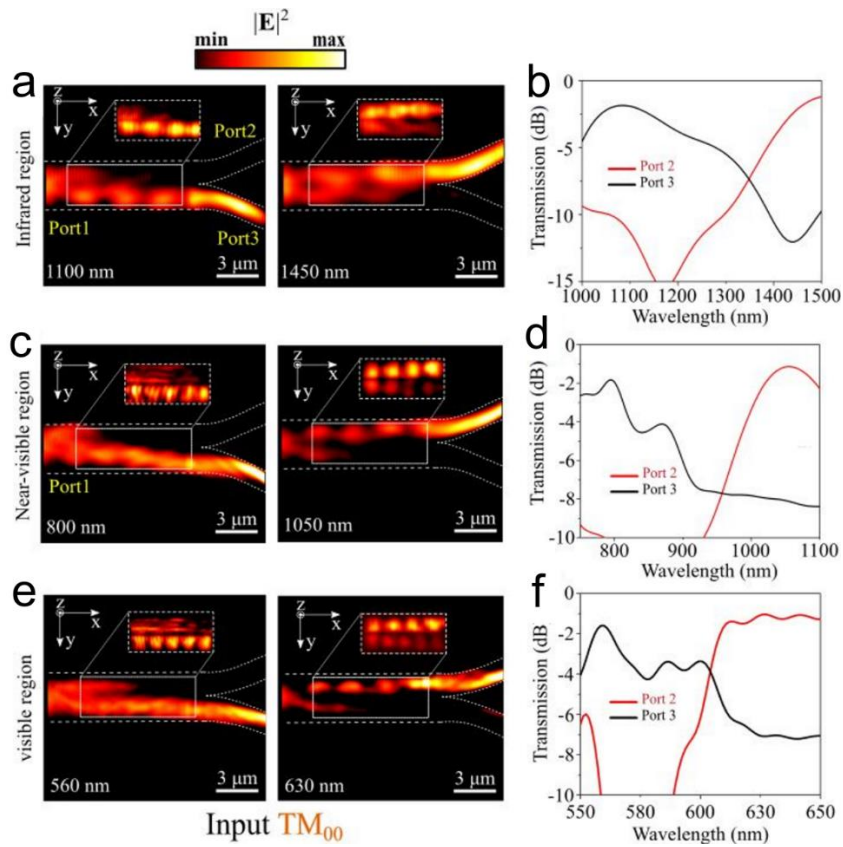

**Figure 4:** WDDs performances for input TM$_{00}$ mode. **a, c, e)** The simulated field intensity profiles $|E|^2$ along the proposed WDDs at the (XY) plane for the fundamental TM$_{00}$ mode operating in the infrared, near-visible, and visible regions, respectively. The dashed lines and rectangles in the figures show the boundaries of the input/output SiN ports and the superimposed TiO$_2$ metasurface. **b, d, f)** The calculated transmission at the output ports of the corresponding devices over the simulated wavelengths. The pass-bands are clearly distinguished in these figures. The observed peaks of transmission are at 1100 nm, 1450

nm, 850, 1050 nm, 560 nm and 630 nm, respectively.

1450 nm band, -1.8 dB in the 850 nm band, -1 dB in the 1050 nm band, -1.5 dB in the 560 nm band and -1.03 dB in the 630 nm band. The lowest crosstalk in all the bands is ~ -10.1 dB. We note that our design approach can also be applied for devices working with polarized-$TM_{00}$ light. To ensure that the input $TE_{00}/TM_{00}$ modes maintain their polarization states after the interactions with the metasurfaces, the electric field components at the output ports of the devices are plotted in **Figs. S4.**

### 1.3.2 Three-channel WDDs

Our design approach can also be applied for realizing multi-channel WDDs. **Figure 5** shows the functionality and schematic layout of the proposed three-channel WDDs. The device consists of three nanoantenna-arrays of $TiO_2$ located on the top center of the SiN waveguide that has a width ($d$=5 µm), a ridge height ($h$=220 nm), and under-etched-film height ($s$=200 nm). The output branches (output ports) have a width of (1.6 µm). The a-$TiO_2$ nanoantennas have a rectangular cross-section with identical length ($L$=1.1 µm), height ($h$=250 nm), and footprint ($L_x$=10 µm). The arrays are separated by a gap ($g$=180 nm) and oriented along the x-axis. Each antenna-array consists of n nanorods, width $W_n$, and pitch period $\Lambda_n$. **Table S2** provides detailed parameters for the optimized nanoantenna-arrays of the proposed three-channel WDDs devices.

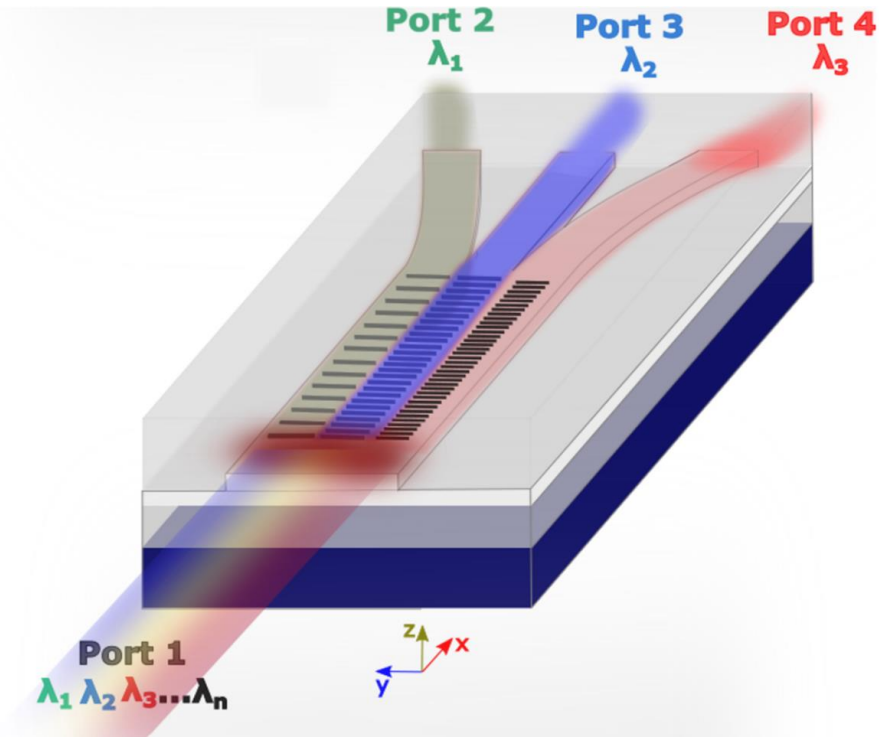

**Figure 5:** Design of Three-channel WDDs. **a)** The schematic illustration of the proposed metasurface-based integrated 3-channel WDDs.

**Figures 6a, and Fig. 6c** show the simulated field intensity profiles $|E|^2$ for input a) $TE_{00}$ and b) $TE_{00}$ mode along the propagation direction (x-axis) of the device at different pass-bands in 1500 nm (left), 1050 nm (middle), and 900 nm (right). The simulated transmission spectra at output ports (port2, port3, port4) of the corresponding devices are plotted in **Figs. 6b, and Fig. 6d**. For WDDs designed for $TE_{00}$ modes, the peak transmission is -3.4 dB in 1500 nm band, -1.6 dB in 1050 nm band, -3 dB in 900 nm band, and -1 dB in 1050 nm band. For the $TM_{00}$ modes, the peak transmission is -3.4 dB in the 1500 nm band, -1.6 dB in the 1050 nm band, and -3 dB in the 900 nm band. The simulated crosstalk at peak transmission for $TE_{00}$ and $TM_{00}$ was -12 dB and -8 dB, respectively. Note that the input $TE_{00}/TM_{00}$ modes maintain their polarization states after the interactions with the metasurfaces, as indicated by the electric field components at the output ports of the devices in **Figs. S5**.

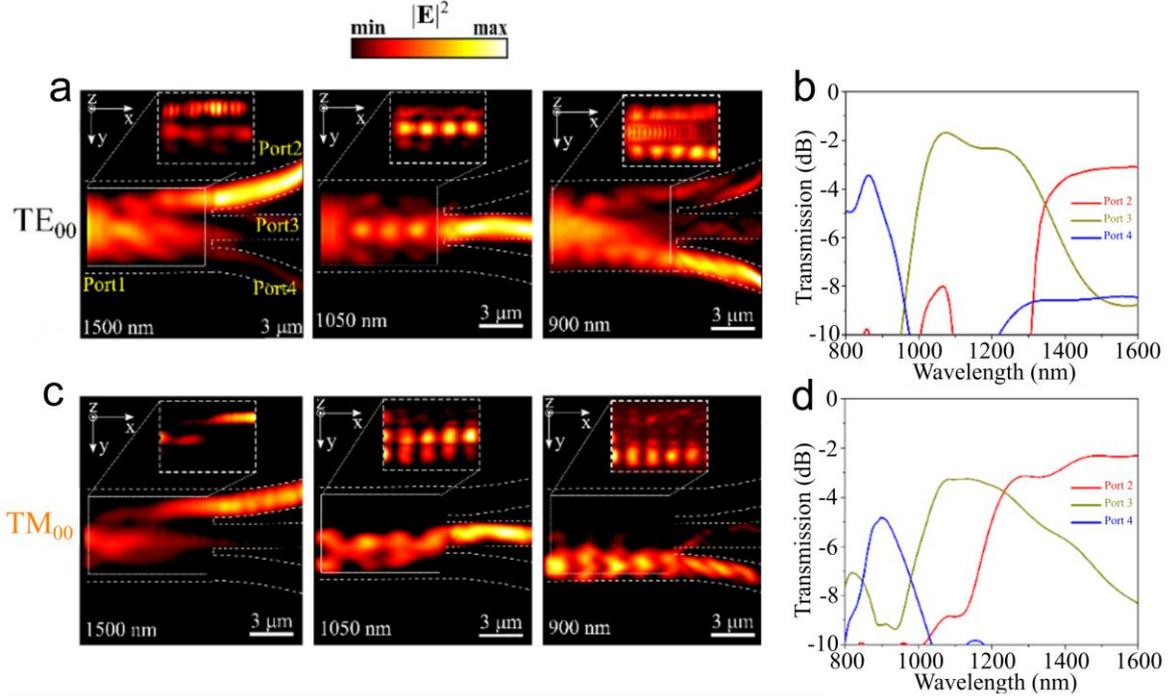

**Figure 6**: WDDs performances for input $TE_{00}$ and $TM_{00}$ modes. **a, c)** The simulated field intensity profiles $|E|^2$ along the proposed WDDs at the (XY) plane operating at three distinct wavelengths for the fundamental $TE_{00}$ and $TM_{00}$ mode, respectively. The dashed lines and rectangles in the figures show the boundaries of the input/output SiN ports and the superimposed $TiO_2$ metasurface. **b, d)** The calculated transmission at the output ports of the corresponding devices over the simulated wavelengths. The three pass-bands around (900 nm, 1050 nm, 1500 nm) are clearly distinguished in the figures.

## Conclusion

In conclusion, we presented a design approach for realizing wavelength-division demultiplexing schemes through integrating arrays of $TiO_2$ nanoantennas onto SiN waveguides. The design concept is based on the wavelength-dependent local effective index engineering of SiN index induced by $TiO_2$ meta-atoms arrays. We simulated 2-channel and 3-channel WDDs operating for $TE_{00}$ and $TM_{00}$ modes in different wavelength bands, proving the robustness and flexibility of our design approach. The proposed WDDS are compatible with photolithography-based processes and display a high performance (maximum transmission 91%) and broadband operations with large channel spacing. This work motivates realizing other compact metasurface-based photonic integrated interconnects for applications in high-bit-rate optical data processing and biological sensing.

# Acknowledgment

A. A. acknowledges the CAS-TWAS presidents fellowship program. National Natural Science Foundation of China (grant nos. 62134009, 62121005)

# References

1. Bogaerts, W., et al., *Programmable photonic circuits.* Nature, 2020. **586**(7828): p. 207-216.
2. Bogaerts, W. and L. Chrostowski, *Silicon photonics circuit design: methods, tools and challenges.* Laser & Photonics Reviews, 2018. **12**(4): p. 1700237.
3. Koos, C., et al., *All-optical high-speed signal processing with silicon–organic hybrid slot waveguides.* Nature Photonics, 2009. **3**(4): p. 216-219.
4. Athale, R. and D. Psaltis, *Optical Computing: Past and Future.* Optics and Photonics News, 2016. **27**(6): p. 32-39.
5. Ren, M., et al., *Nanostructured Plasmonic Medium for Terahertz Bandwidth All-Optical Switching.* Advanced Materials, 2011. **23**(46): p. 5540-5544.
6. Dai, D. and J.E. Bowers, *Silicon-based on-chip multiplexing technologies and devices for Peta-bit optical interconnects.* Nanophotonics, 2014. **3**(4-5): p. 283-311.
7. Wiecha, P.R., et al., *Deep learning in nano-photonics: inverse design and beyond.* Photonics Research, 2021. **9**(5): p. B182-B200.
8. Zhang, W., et al., *Designing crystallization in phase-change materials for universal memory and neuro-inspired computing.* Nature Reviews Materials, 2019. **4**(3): p. 150-168.
9. Shastri, B.J., et al., *Photonics for artificial intelligence and neuromorphic computing.* Nature Photonics, 2021. **15**(2): p. 102-114.
10. Huang, J., et al., *Digital nanophotonics: the highway to the integration of subwavelength-scale photonics.* Nanophotonics, 2021. **10**(3): p. 1011-1030.
11. Larsen, M.V., et al., *Deterministic multi-mode gates on a scalable photonic quantum computing platform.* Nature Physics, 2021.
12. MacQuarrie, E.R., et al., *The emerging commercial landscape of quantum computing.* Nature Reviews Physics, 2020. **2**(11): p. 596-598.
13. Marpaung, D., J. Yao, and J. Capmany, *Integrated microwave photonics.* Nature Photonics, 2019. **13**(2): p. 80-90.
14. Marpaung, D., et al., *Integrated microwave photonics.* Laser & Photonics Reviews, 2013. **7**(4): p. 506-538.
15. Piggott, A.Y., et al., *Inverse design and demonstration of a compact and broadband on-chip wavelength demultiplexer.* Nature Photonics, 2015. **9**(6): p. 374-377.
16. Piggott, A.Y., et al., *Inverse-Designed Photonics for Semiconductor Foundries.* ACS Photonics, 2020. **7**(3): p. 569-575.
17. Luo, L.-W., et al., *WDM-compatible mode-division multiplexing on a silicon chip.* Nature Communications, 2014. **5**(1): p. 3069.
18. Mori, Y., et al., *Wavelength-Division Demultiplexing Enhanced by Silicon-Photonic Tunable Filters in Ultra-Wideband Optical-Path Networks.* Journal of Lightwave Technology, 2020. **38**(5): p. 1002-1009.

19. Feng, C., et al., *Wavelength-division-multiplexing (WDM)-based integrated electronic–photonic switching network (EPSN) for high-speed data processing and transportation.* Nanophotonics, 2020. **9**(15): p. 4579-4588.
20. Horst, F., et al., *Silicon-on-Insulator Echelle Grating WDM Demultiplexers With Two Stigmatic Points.* IEEE Photonics Technology Letters, 2009. **21**(23): p. 1743-1745.
21. Sciancalepore, C., et al., *Low-Crosstalk Fabrication-Insensitive Echelle Grating Demultiplexers on Silicon-on-Insulator.* IEEE Photonics Technology Letters, 2015. **27**(5): p. 494-497.
22. van Wijk, A., et al., *Compact ultrabroad-bandwidth cascaded arrayed waveguide gratings.* Optics Express, 2020. **28**(10): p. 14618-14626.
23. Rank, E.A., et al., *Toward optical coherence tomography on a chip: in vivo three-dimensional human retinal imaging using photonic integrated circuit-based arrayed waveguide gratings.* Light: Science & Applications, 2021. **10**(1): p. 6.
24. Chen, Y., et al., *Uniform-loss cyclic arrayed waveguide grating router using a mode-field converter based on a slab coupler and auxiliary waveguides.* Optics Letters, 2019. **44**(2): p. 211-214.
25. Dahlem, M.S., et al., *Reconfigurable multi-channel second-order silicon microring-resonator filterbanks for on-chip WDM systems.* Optics Express, 2011. **19**(1): p. 306-316.
26. Shen, H., et al., *Eight-channel reconfigurable microring filters with tunable frequency, extinction ratio and bandwidth.* Optics Express, 2010. **18**(17): p. 18067-18076.
27. Nawwar, O.M., H.M.H. Shalaby, and R.K. Pokharel, *Photonic crystal-based compact hybrid WDM/MDM (De)multiplexer for SOI platforms.* Optics Letters, 2018. **43**(17): p. 4176-4179.
28. Xu, L., et al., *Broadband 1310/1550 nm wavelength demultiplexer based on a multimode interference coupler with tapered internal photonic crystal for the silicon-on-insulator platform.* Optics Letters, 2019. **44**(7): p. 1770-1773.
29. Mikkelsen, J.C., et al., *Polarization-insensitive silicon nitride Mach-Zehnder lattice wavelength demultiplexers for CWDM in the O-band.* Optics Express, 2018. **26**(23): p. 30076-30084.
30. Yen, T.-H. and Y. Hung, Jr., *Fabrication-Tolerant CWDM (de)Multiplexer Based on Cascaded Mach–Zehnder Interferometers on Silicon-on-Insulator.* Journal of Lightwave Technology, 2021. **39**(1): p. 146-153.
31. Rouifed, M.S., et al., *Ultra-compact MMI-based beam splitter demultiplexer for the NIR/MIR wavelengths of 1.55 µm and 2 µm.* Optics Express, 2017. **25**(10): p. 10893-10900.
32. Xiao, J., X. Liu, and X. Sun, *Design of an ultracompact MMI wavelength demultiplexer in slot waveguide structures.* Optics Express, 2007. **15**(13): p. 8300-8308.
33. Liu, J.S.Q., et al., *A submicron plasmonic dichroic splitter.* Nature Communications, 2011. **2**(1): p. 525.
34. Aparna, U., R. Mendiratta, and L.K. Shrinidhi, *1 × 2 plasmonic wavelength demultiplexer using rectangular MIM waveguide.* Journal of Optical Communications, 2020.
35. Wu, C.-T., C.-C. Huang, and Y.-C. Lee, *Plasmonic wavelength demultiplexer with a ring resonator using high-order resonant modes.* Applied Optics, 2017. **56**(14): p. 4039-4044.
36. Liu, H., et al., *A T-shaped high resolution plasmonic demultiplexer based on perturbations of two nanoresonators.* Optics Communications, 2015. **334**: p. 164-169.
37. Fan, Y., et al., *Ultra-compact on-chip metaline-based 1.3/1.6 µm wavelength demultiplexer.* Photonics Research, 2019. **7**(3): p. 359-362.
38. Foresi, J.S., et al., *Photonic-bandgap microcavities in optical waveguides.* Nature, 1997. **390**(6656): p. 143-145.
39. Lu, C., et al., *Integrated ultracompact and broadband wavelength demultiplexer based on multi-component nano-cavities.* Scientific Reports, 2016. **6**(1): p. 27428.

40. Su, L., et al., *Inverse Design and Demonstration of a Compact on-Chip Narrowband Three-Channel Wavelength Demultiplexer.* ACS Photonics, 2018. **5**(2): p. 301-305.
41. Huang, J., et al., *Implementation of on-chip multi-channel focusing wavelength demultiplexer with regularized digital metamaterials.* Nanophotonics, 2020. **9**(1): p. 159-166.
42. Ma, W., et al., *Deep learning for the design of photonic structures.* Nature Photonics, 2021. **15**(2): p. 77-90.
43. Shen, Y., et al., *Deep learning with coherent nanophotonic circuits.* Nature Photonics, 2017. **11**(7): p. 441-446.
44. Piggott, A.Y., et al., *Fabrication-constrained nanophotonic inverse design.* Scientific Reports, 2017. **7**(1): p. 1786.
45. Liu, Z., et al., *Integrated nanophotonic wavelength router based on an intelligent algorithm.* Optica, 2019. **6**(10): p. 1367-1373.
46. Staude, I., T. Pertsch, and Y.S. Kivshar, *All-Dielectric Resonant Meta-Optics Lightens up.* ACS Photonics, 2019. **6**(4): p. 802-814.
47. Ohana, D., et al., *Dielectric Metasurface as a Platform for Spatial Mode Conversion in Nanoscale Waveguides.* Nano Letters, 2016. **16**(12): p. 7956-7961.
48. Koshelev, K. and Y. Kivshar, *Dielectric Resonant Metaphotonics.* ACS Photonics, 2021. **8**(1): p. 102-112.
49. Chen, Z. and M. Segev, *Highlighting photonics: looking into the next decade.* eLight, 2021. **1**(1): p. 2.
50. Scheuer, J., *Optical Metasurfaces Are Coming of Age: Short- and Long-Term Opportunities for Commercial Applications.* ACS Photonics, 2020. **7**(6): p. 1323-1354.
51. Shaltout, A.M., V.M. Shalaev, and M.L. Brongersma, *Spatiotemporal light control with active metasurfaces.* Science, 2019. **364**(6441): p. eaat3100.
52. Choudhury, S.M., et al., *Material platforms for optical metasurfaces.* Nanophotonics, 2018. **7**(6): p. 959-987.
53. Yu, N. and F. Capasso, *Flat optics with designer metasurfaces.* Nature Materials, 2014. **13**(2): p. 139-150.
54. Overvig, A.C., et al., *Dielectric metasurfaces for complete and independent control of the optical amplitude and phase.* Light: Science & Applications, 2019. **8**(1): p. 92.
55. Hu, Y., et al., *All-dielectric metasurfaces for polarization manipulation: principles and emerging applications.* Nanophotonics, 2020. **9**(12): p. 3755-3780.
56. Li, Z., et al., *Controlling propagation and coupling of waveguide modes using phase-gradient metasurfaces.* Nature Nanotechnology, 2017. **12**(7): p. 675-683.
57. Wang, C., et al., *Metasurface-assisted phase-matching-free second harmonic generation in lithium niobate waveguides.* Nature Communications, 2017. **8**(1): p. 2098.
58. Zhang, Y., et al., *Spin-Selective and Wavelength-Selective Demultiplexing Based on Waveguide-Integrated All-Dielectric Metasurfaces.* Advanced Optical Materials, 2019. **7**(6): p. 1801273.
59. Alquliah, A., et al., *Ultrabroadband, compact, polarization independent and efficient metasurface-based power splitter on lithium niobate waveguides.* Optics Express, 2021. **29**(6): p. 8160-8170.
60. Meng, Y., et al., *Optical meta-waveguides for integrated photonics and beyond.* Light: Science & Applications, 2021. **10**(1): p. 235.
61. Wang, Z., et al., *Metasurface on integrated photonic platform: from mode converters to machine learning.* Nanophotonics, 2022.
62. Lian, C., et al., *Photonic (computational) memories: tunable nanophotonics for data storage and computing.* Nanophotonics, 2022.
63. Blumenthal, D.J., et al., *Silicon Nitride in Silicon Photonics.* Proceedings of the IEEE, 2018.

**106**(12): p. 2209-2231.
64. Baets, R., et al. *Silicon Photonics: silicon nitride versus silicon-on-insulator*. in *Optical Fiber Communication Conference*. 2016. Anaheim, California: Optical Society of America.
65. Khorasaninejad, M., et al., *Visible Wavelength Planar Metalenses Based on Titanium Dioxide.* IEEE Journal of Selected Topics in Quantum Electronics, 2017. **23**(3): p. 43-58.
66. Wu, Y., et al., *TiO2 metasurfaces: From visible planar photonics to photochemistry.* Science Advances, 2019. **5**(11): p. eaax0939.
67. Palik, E.D., *Handbook of optical constants of solids*. Vol. 2. 2012: Academic press.
68. Cheben, P., et al., *Subwavelength integrated photonics.* Nature, 2018. **560**(7720): p. 565-572.
69. Sun, L., et al., *Subwavelength structured silicon waveguides and photonic devices.* 2020. **9**(6): p. 1321-1340.
70. Halir, R., et al., *Waveguide sub-wavelength structures: a review of principles and applications.* Laser & Photonics Reviews, 2015. **9**(1): p. 25-49.
71. Guo, X., et al., *Molding free-space light with guided wave–driven metasurfaces.* Science Advances, 2020. **6**(29): p. eabb4142.
72. Alquliah, A., et al., *Reconfigurable metasurface-based 1 x 2 waveguide switch.* Photonics Research, 2021. **9**(10): p. 2104-2115.
73. Sun, L., et al., *Subwavelength structured silicon waveguides and photonic devices.* Nanophotonics, 2020. **9**(6): p. 1321-1340.
74. Rytov, S., *Electromagnetic properties of a finely stratified medium.* Soviet Physics Journal of Experimental and Theoretical Physics, 1956. **2**: p. 466-475.